
\magnification=\magstep1
\hsize=6.3truein
\vsize=8.375truein
\parindent=20pt
\nopagenumbers
\raggedbottom
\hangafter=1
\def\makeheadline{\vbox to 0pt{\vskip-40pt
   \line{\vbox to 8.5pt{}\the\headline}\vss}\nointerlineskip}
\def\approxlt{\kern 0.35em\raise 0.45ex\hbox{$<$}\kern-0.66em\lower0.5ex
   \hbox{$\scriptstyle\sim$}\kern0.35em}
\def\approxgt{\kern 0.35em\raise 0.45ex\hbox{$>$}\kern-0.75em\lower0.5ex
   \hbox{$\scriptstyle\sim$}\kern0.35em}
\vglue1.5truein
\centerline{\bf Applications of W Algebras to BF Theories, QCD, and
4D Gravity}
\vskip22pt
\centerline{R.P. Lano and V.G.J. Rodgers}
\centerline{ Department of Physics and Astronomy}
\centerline{ The University of Iowa}
\centerline{ Iowa City, Iowa~~52242--1479}
\centerline{ March 1992 }
\vglue0.5truein
\centerline{Dedicated to Ms. Miya Sonya Sioson}
\centerline{and to the memory of}
\centerline{ Prof. Dwight Nicholson,
Prof. Bob Smith, Prof. Chris Goertz,}
\centerline{ Dr. Anne Cleary and
Dr. Linhua Shan}
\vglue0.5truein
\baselineskip=12pt
\centerline{\bf ABSTRACT}
\vskip22pt
We are able to show that BF theories naturally emerge from the
coadjoint orbits of $W_2$ and $w_\infty$ algebras which includes
a Kac-Moody sector.  Since QCD strings can be identified with
a BF theory, we are able to show a relationship between the
orbits and monopole-string solutions of QCD.  Furthermore,
we observe that when 4D gravitation is cast into a BF form
through the use of Ashtekar variables, we are able to get
order $\hbar$ contributions to gravity which can be associated
with the $W_2$ anomaly.  We comment on the relationship to
gravitational monopoles.
\vfill
\eject
\headline={\tenrm\hfil\folio}
\baselineskip=16pt
\pageno=1
Topological field theories are today considered an indispensable
tool in the zeroth and first order attempts to understand many
physical systems and mathematical issues.  In particular
BF and Chern-Simons theories have provided physicists with
an enormous amount of insight into what we should look for in systems
such as
3d quantum gravity [1], 4D gravity [2-3],
superconductivity, and the London equations [4].
In this letter we would like to contribute to this wealth of
knowledge by addressing the question of BF anomalies coming from the
$w_\infty$ and $W_2$ algebras.  We will relate these anomalies
to QCD and 4D gravity.  The methods we use are well known [5-10] and rely
heavily on the existence of a symplectic two form for the infinite
dimensional algebra.  Therefore other $W_n$ algebras cannot be considered
from this viewpoint.   Our first calculations will be with respect
to a three dimensional BF theory.  Then we construct a four dimensional
BF theory that enjoys $w_\infty$ symmetry.  We will arrive at the BF theories
``through the back door.''  In other words our approach is to ask,
``given a specific symmetry, what action yields the associated
Noether currents.''
\bigskip

{\bf Three Dimensional Systems}
\vskip2pt
Let us begin by studying a Kac-Moody algebra and the diffeomorphism
of the circle parameter [11].  One recalls that there
we have a Kac-Moody algebra associated with the group G given by
$$
\eqalign{
 [ J^{\alpha}_N, J^{\beta}_M] & = i f^{\alpha\beta\gamma} J^\gamma_{N+M} + N k
\delta_{M+N,0} \delta^{\alpha\beta} \cr
[ L_N,J^{\alpha}_M] & = - M J^{\alpha}_{M+N} \cr
[ L_N,L_M ] &= (N-M) L_{N+M} + {c\over 12} (N^3-N) \delta_{N+M,0}}
$$
where $c = {2 k Dim(G)\over 2k + c_v}$, $Dim(G)$ is the dimension
of the group and $c_v$ is the value of the quadratic Casimir
in the adjoint representation.  Now the choice of representations
depends on the topology of the manifold.   Let us choose the
representation,
$$ \eqalign{ L_N& = i \exp^{i N\theta} \partial_\theta \cr
J^\alpha_N& = \tau^\alpha \exp^{i N \theta}.}
$$
We normalize the generators so that $Tr( \tau^\alpha
\tau^\beta)=\delta^{\alpha\beta}$.
Let us denote $\bigl( L_A, J^\beta_B, \rho\bigr)$ as a centrally extended
adjoint
vector.  Then from the commutation relations above one may write the
adjoint action on the adjoint vectors as
$$\eqalign{
\bigl(& L_A, J^\beta_B, \rho\bigr) \ast \bigl( L_{N'}, J^{\alpha
'}_{M'},\mu\bigr) = \cr
&\bigl( (A-N') L_{A+N'},\cr
& -M' J^{\alpha '}_{A+M'} +B J^\beta_{B+N'} +i f^{\beta \alpha ' \lambda}
J^\lambda_{B+M'} ,\cr
& {c\over 12} (A^3-A) \delta_{A+N',0} + B k \delta^{\alpha '\beta}
\delta_{B+M',0}\bigr).}
$$
Now we need the coadjoint representation and a suitable inner product.
Let $\bigl( {\tilde L_N}, {\tilde J^\alpha_M}, {\tilde \mu} \bigr)$
denote an element of the coadjoint representation.  A suitable
pairing is simple
$$\bigl\langle\bigl( {\tilde L_N}, {\tilde J^\alpha_M}, {\tilde \mu} \bigr)
\bigr\vert
   \bigl( L_{N'}, J^{\alpha '}_{M'}, \mu ' \bigr)\bigr\rangle = \delta^{N,N'} +
\delta^{\alpha,\alpha '} \delta_{M, M'} + \mu {\tilde \mu}. $$
With this the adjoint action on the coadjoint representation
can be extracted by requiring that
$$\eqalign{
&\bigl\langle \bigl( {\tilde L_N}, {\tilde J^\alpha_M}, {\tilde \mu} \bigr)
\bigr\vert
\bigl( L_A, J^\beta_B, \rho\bigr) \ast \bigl( L_{N'}, J^{\alpha
'}_{M'},\mu\bigr) \bigr\rangle =\cr
 - &\bigl\langle \bigl( L_A, J^\beta_B, \rho\bigr) \ast \bigl( {\tilde L_N},
{\tilde J^\alpha_M}, {\tilde \mu} \bigr) \bigr\vert \bigl( L_{N'}, J^{\alpha
'}_{M'},\mu\bigr) \bigr\rangle .}
$$
This means that the action of an adjoint vector on a coadjoint vector
yields a new coadjoint vector through,
$$\eqalign{
& \bigl( L_A, J^\beta_B, \rho\bigr) \ast \bigl( {\tilde L_N}, {\tilde
J^\alpha_M}, {\tilde \mu}\bigr)  = \cr
& \bigl( -(2A-N) {\tilde L_{N-A}} -B \delta^{\alpha\beta} {\tilde L_{M-B}}
-{{ \tilde \mu} c\over 12} (A^3-A) {\tilde L_{-A}},\cr
&
(M-A) {\tilde J^\alpha_{M-A}} -i f^{\beta\nu\alpha} {\tilde J^{\nu}_{M-B}} -
{\tilde \mu} B k {\tilde J^\beta_{-B}}, 0 \bigr).} \eqno(1)$$
By setting this equation to zero one can  determine the isotropy algebra for
the covector $\bigl( {\tilde L_N}, {\tilde J^\alpha_M}, {\tilde \mu}\bigr)$.
Instead of using components, let us write $F = (f(\theta),{\hat h}(\theta),a)$
as an arbitrary adjoint vector and $B=(b(\theta),h(\theta),\mu)$ as an
arbitrary coadjoint vector, where $f, {\hat h}, b,$ and $h $ are functions
constructed from the above mentioned basis vectors.  Then Eq.~(1) may
be written as
$$\eqalign{
&\delta_F B \equiv (f(\theta),{\hat h}(\theta),a) \ast
(b(\theta),h(\theta),\mu) = \cr
-&\bigl( 2 f'b+b'f+i  {c \mu\over 12} f''' + Tr[ h {\hat h}'],
 h'f+hf' + [{\hat h} h - h {\hat h}] + i k \mu {\hat h}', 0 \bigr),}\eqno(1a)
$$
where $'$ denotes $\partial_\theta$.
Those adjoint vectors, $F$, that leave $B$ invariant will generate the
isotropy group for $B$.  The group action of G on $B$ is generated
by the adjoint representation for those elements of the group that
are connected to the identity.  Equation (1a) then determines the tangent
space on the orbit of $B$.  Thus for coadjoint elements $B_1$ and $B_2$,
we may construct the usual symplectic
two form by writing
$$
\Omega_B(B_1,B_2) = \langle B \mid [ F_1,F_2] \rangle,
$$ where for example $\delta_{F_1} B = B_1$.

Next we will need the two cocycle associated with the central
extension for our representation.  For adjoint elements
$(L,J,\mu)$ and $({\hat L},{\hat J}, {\hat \mu})$ we may define
the two cocycle as,
$$
\eqalign{
& C_{\rm Diff S^1\times SU(N)}\bigl((L,J,\mu), ({\hat L},{\hat J}, {\hat
\mu})\bigr) =\cr
{i c \over 48 \pi} &\int^{2\pi}_0 ( L(\theta) {\hat L(\theta)}'''-
 {\hat L(\theta)} L(\theta)''' ) d\theta  \cr
+ {k\over2\pi} &\int^{2\pi}_0 Tr J(\theta) \partial_\theta {\hat J(\theta)}
d\theta.}
$$
Now the idea of constructing actions from coadjoint orbits is very simple.
One simply considers the orbit of $B$ as a manifold.  Since G action on
$B$ corresponds to a new point on the orbit, we may think of G modulo
elements of the isotropy group of $B$ as parameterizing the orbit.
Then consider a two parameter family of group elements living on the
orbit.  These two parameters can be associated with a dynamical variable
such as time, $\tau$ and a one parameter ($\lambda$)  family of maps from the
identity (say)
to a fixed element ${\tilde g(\lambda=1,\tau)}$.  Thus our group elements may
be written as
${\tilde g}(\lambda, \tau)$ and these correspond to generic Kac-Moody group and
diffeomorphism group elements.  We need to know the tangent covectors
associated
with infinitesimal transformations along the flow of $\lambda$ and $\tau$.
Then one may simply write that
$$
S_B = \int d\lambda d\tau ~ \Omega \big( B^{{\tilde g}(\lambda,\tau)}_\lambda,
B^{{\tilde g}(\lambda,\tau)}_\tau \big) =
 \int d\lambda d\tau \bigl\langle B_{{\tilde g}(\lambda,\tau)} \bigr\vert \big[
F^{{\tilde g}(\lambda,\tau)}_\lambda, F^{{\tilde g}(\lambda,\tau)}_\tau \big]
\bigr\rangle.
$$
Here $F^{{\tilde g}(\lambda,\tau)}_\lambda$ and $F^{{\tilde
g}(\lambda,\tau)}_\tau$ are the directional derivatives (adjoint vectors)
associated with flow along $\lambda$ and $\tau$ respectively at the point
determined by ${\tilde g}(\lambda,\tau)$.
We simply need to disentangle this expression for our purposes.
Because of the central extensions, the coadjoint action of the group will
in general have  an inhomogeneous term.
One can show that [9] in these cases the action is
$$
S=\int d\lambda\,d\tau\biggl(\langle  b \mid [{\tilde g}^{-1}F^{\tilde
g}_\lambda {\tilde g} ,{\tilde g} ^{-1} F^{{\tilde g}}_\tau g]\rangle
+ \mu c({\tilde g}^{-1} F^{{\tilde g}}_{\lambda} {\tilde g},{\tilde
g}^{-1}F^{{\tilde g}}_{\tau}{\tilde g})\biggr)\ , \eqno(2)$$
where the centrally extended coadjoint vector in consideration here is
$B=(b,\mu)$, with $b$ representing all classical components and $\mu$
representing the central extensions.  In the above ${\tilde g}^{-1} F^{{\tilde
g}}_{\tau}{\tilde g}$ represents the
pull back of the adjoint vector to the identity and $b$ is evaluated at
the identity.
With this we find that
$$\eqalign{
F^{\tilde g}_\lambda &= ( \partial_\lambda s(\theta, \lambda, \tau),
\partial_\lambda g g^{-1},0)\cr
F^{\tilde g}_\tau & = ( \partial_\tau s(\theta, \lambda, \tau), \partial_\tau g
g^{-1},0)} \eqno(3)
$$
and
$$\eqalign{
{\tilde g}^{-1}F^{\tilde g}_\lambda {\tilde g}& = \bigg( {\partial_\lambda
s(\theta, \lambda, \tau)\over \partial_\theta s }, g^{-1} \partial_\lambda g ,0
\bigg)\cr
{\tilde g}^{-1}F^{\tilde g}_\tau {\tilde g}& = \bigg( {\partial_\tau s(\theta,
\lambda, \tau)\over \partial_\theta s }, g^{-1} \partial_\tau g ,0 \bigg)}
\eqno(3a)
$$
where $s(\theta,\lambda,\tau)$ corresponds to a two parameter family of
diffeomorphisms of the circle parameter and $g(\theta,\lambda,\tau)$
is a two parameter family of Kac-Moody group elements.
With all of this in place we arrive at our result which corresponds to
an action that is both Kac-Moody and diffeomorphism invariant.

$$\eqalign{
&S = \int b(\theta) \bigg\lbrack {\partial_\lambda s \over \partial_\theta s}
{\partial \over \partial \theta} \bigg( { \partial_\tau s \over \partial_\theta
s}\bigg)
-  {\partial_\tau s \over \partial_\theta s} {\partial \over \partial \theta}
\bigg( { \partial_\lambda s \over \partial_\theta s} \bigg)
\bigg\rbrack~d\lambda d\tau d\theta
\cr
&+{k\over 2\pi} \int Tr \biggl[h(\theta)\bigg\lbrace
 {\partial_\lambda s \over \partial_\theta s}{\partial \over \partial \theta}
(g^{-1}\partial_\tau g)-{\partial_\tau s \over \partial_\theta s}{\partial
\over \partial \theta} (g^{-1}\partial_\lambda g)
+[ g^{-1}\partial_\lambda g,
g^{-1}\partial_\tau g] \bigg\rbrace \biggr]d\lambda d\tau d\theta
\cr
&+{c(k,c_v,26) \mu  \over 48\pi } \int  \left[
{{\partial^2_{\theta} s}\over{(\partial_{\theta}s)^2}} \partial
_{\tau} \partial_{\theta} s -  {{(\partial^2_{\theta}s)^2
(\partial_{\tau} s)}\over{(\partial_{\theta} s)^3}} \right] d\theta d\tau
     - {{n \mu}\over{4 \pi}} \int Tr g^{-1} {{\partial g}\over{\partial
\theta}}g^{-1}
{{\partial g}\over{\partial \tau}}d \theta d \tau
\cr
&+{{k \mu}\over{4 \pi}} \int Tr g^{-1} {{\partial g}\over{\partial \theta}}
\left[ g^{-1}
{{\partial g}\over{\partial \lambda}}, g^{-1} {{\partial g}\over{\partial
\tau}} \right]
d\theta d\tau d\lambda.} \eqno(4)
$$
In the above action we have written $c(k,c_v,26)$ as the central extension
shifted by 26 in accord with reparameterization ghosts that arise from
a covariant treatment.  The covector defining the orbit is
$B = (b(\theta),h(\theta),\mu)$.  By setting Eq.~(1a) to zero one can
recover the isotropy group for $B$.  Keep in mind that $b(\theta)$ is
a quadratic differential and the $h(\theta)$ is valued in the Lie
algebra of G.

Now the fact that this is a BF theory in three dimensions can be seen from the
first two lines of Eq.~(4).  Indeed by writing the Kac-Moody $\times$ Diff
$S^1$
valued curvature two forms as
$$\eqalign{
{\bf F}_{\lambda\tau} =
& {\partial_\lambda s \over \partial_\theta s}{\partial \over \partial \theta}
(g^{-1}\partial_\tau g)-{\partial_\tau s \over \partial_\theta s}{\partial
\over \partial \theta} (g^{-1}\partial_\lambda g)
+[ g^{-1}\partial_\lambda g,
g^{-1}\partial_\tau g] \cr
\noalign{\vskip5pt}
+&\biggl( {\partial_\lambda s \over \partial_\theta s} {\partial \over \partial
\theta} \bigg( { \partial_\tau s \over \partial_\theta s}\bigg)
-  {\partial_\tau s \over \partial_\theta s} {\partial \over \partial
\theta}\bigg( { \partial_\lambda s \over \partial_\theta s}\bigg)\biggr)}
\eqno(5)
$$
and writing the corresponding one form {\bf B} as
$
{\bf B}_\theta = h(\theta) + b(\theta),
$ and all other terms zero, we see that the above action is just
$$ S = \int Tr (\epsilon^{ijk} B_i F_{jk}) d\lambda d\tau d\theta = \int Tr
B_\wedge F.
$$
Note that the last term of {\bf F} and the $b(\theta)$ term of {\bf B}
are related to an abelian subalgebra.  Let us remark that throughout we have
assumed a three dimensional topology of $S^1\times R^2$.  If one wishes to
change the topology, this will modify the central extension term.  However
the analysis would be straightforward.  Now the last two summands of Eq.~(4)
are order $\hbar$ contributions to the theory.  Since a BF theory in
three dimensions may be written as a Chern-Simons theory [1,12]
we find that a WZNW model and a Liouville-Polyakov model appear at the
first level of quantum corrections.  In fact these terms may be viewed
as the effective action of a BF theory or the bosonization of a
fermionic theory coupled to an external gauge potential and 2D gravity.
Let us discuss some of the applications of this model.

In Ref.~[13] the author has argued that the vacuum sector for
QCD may best be described as a  WZNW model immersed into a four dimensional
manifold.  By considering an SU(3) Kac-Moody group we can recover that
string theory modulo self-intersection terms.  We observe that the
gravity from the string world surface introduces an anomaly
via the Liouville-Polyakov term.  This term will break the scale
invariance and may possibly be related to $\Lambda_{QCD}$.
It is plausible that
since the two dimensional gravity associated with the string world surface
arises from the induced metric of four dimensions that the scale
anomaly of QCD is due to gravity.  This would imply that the
dilaton arises from quantum gravity effects.   Also the
magnetic-instanton sector [14] of QCD in the string-like phase is directly
related
to the classical BF part of our action.  By choosing
the coadjoint vector ${\bf B} = ( b,h,\mu)$ where $h$ and $b$ are $S^1$
invariant
field configurations (by this we mean that they are $\theta$ independent),
and from Eq. (1a),
$$\eqalign{
&\delta_F B \equiv (f(\theta),{\hat h}(\theta),a) \ast
(b(\theta),h(\theta),\mu) = \cr
-&\bigl( 2 f'b+b'f+i  {c \mu\over 12} f''' + Tr[ h {\hat h}'],
 h'f+hf' + [{\hat h} h - h {\hat h}] + i k \mu {\hat h}', 0 \bigr),}
$$
one sees that the field space is equivalent to $ {{\rm G/H} \otimes \Omega({\rm
G})} \otimes
{{\rm Diff} S^1\over S^1}$, where H is the isotropy group of $g$ and
$\Omega(G)$ is the loop group of G.
The division by $S^1$ is automatic in this case since Eq.~(1a) is still
divisible by rigid rotations from the Diff ${\rm S}^1$ sector.
Note that an $SL(2,C)$ invariant configuration $b(\theta)$ will
spoil the isotropy of $h$.
Different embeddings of H in G will have to be related to
inequivalent self-dual monopoles.
As a final note, the value of $c(k,c_v,26) = 26 - { 8 \times 2k\over
2k + 6}$ (for our normalization of the generators $c_v=6$ as opposed to
3 for the standard normalization $Tr( \tau^a \tau^b) = {1\over2}\delta^{ab}$).
We may also set the value of $k=1$ to yield $c(k,c_v,26) = 24$.

The action also describes a generic G/H coset model [15] coupled to
the Liouville mode.  Observe that in the absence of the Liouville mode
(set $s(\theta,\lambda,\tau) = 0$)  one may integrate the
$h(\theta)[ g^{-1}\partial_\lambda g, g^{-1}\partial_\tau g]$
term by parts [7] to get $h(\theta) g^{-1}\partial_\tau g$.
This is just the light-cone gauge fixed contribution of the G-O
coset models to a WZNW theory.

Since three dimensional gravity is believed to be a BF theory [1]
the action of Eq.~(4) has a further application.  By considering an
ISO(2,1) gauge group and a three manifold with $R^2 \times S^1$
topology, the action above corresponds to a light-cone gauge fixed
BF theory describing a 3D gravitational theory and its order
$\hbar$ corrections.
The Eq.~(5a) and Eq.~(5b) correspond to the curvature contributions
from the frame fields (as functions of $s(\theta)$ and its
derivatives) and the spin connection ($g^{-1}\partial g$).  The
WZNW and Liouville-Polyakov terms serve as the quantum corrections
to this particular 3D topology.  One may consider an $R^3$ topology
by using a different representation for the Virasoro and Kac-Moody generators
such as $L_N=z^{1-N}\partial_z$ and $J^\alpha_{M} = \tau^\alpha z^{-M}$.
In any case the particular choice of the field
${\bf B} = (b(\theta), h(\theta), \mu)$ will choose the coadjoint orbit.
Gravitational monopoles may be related to certain isotropy groups of
{\bf B} just as in the QCD case.
\bigskip
{\bf Four Dimensional Systems}
\vskip2pt
Recently a great deal of research has gone into identifying infinite
dimensional symmetries in self-dual four dimensional gravity [16-19].
In fact $w_\infty$ (for a review see Ref.~[20]) has been identified as a
symmetry in self dual
four dimensional systems through symplectic diffeomorphisms (Diff $M^2$) on two
manifolds
(i.e., they correspond to coordinate transformations of the two dimensional
system
which have unit Jacobian).
At the same time, the work of Ashtekar et al.~[21]
has suggested that a self-dual formulation may be a more
natural framework in which to study quantum gravity in four dimensions.
We would like to use the underlying symplectic geometry of
$w_\infty$ along with a contraction of an SU(N) Kac-Moody algebra in order
to probe self-dual Yang-Mills
systems with these symmetries.  Our hope is to provide some insight
into the quantum corrections to these theories.

Let us begin by writing a suitable representation for $w_\infty$ and
the SU(N) sectors.  The commutations relations are given by
$$\eqalign{
[W^i_M,W^j_N] & = ((j+1)M - (i+1)N) W^{i+j}_{M+N} + {c\over12} (M^3-M)
\delta_{M+N,0} \delta^{i,0}\delta^{j,0}\cr
[J^i_M(\alpha),J^j_N(\beta)] & = i f^{\alpha\beta\gamma}J^{i+j}_{M+N}(\gamma) +
M k \delta_{M+N,0} \delta^{\alpha\beta} \delta^{i,0} \delta^{j,0}\cr
[W^i_N,J^j_M(\alpha)] &= (Nj-Mi-M) J^{i+j}_{M+N}(\alpha),
} \eqno(6)$$
where the central extension $c$ is related to the
Kac-Moody central extension by $c={2k {\rm Dim G} \over 2k+c_v}$.
For the generators $W^i_M$ and $J^i_M(\alpha)$, $i \geq 0, M$ is any
integer and $\alpha = 1,...N^2-1$.
Explicitly the classical part of the generators may be
written as
$$
\eqalign{
W^i_N & = \partial_x(x^{N+1+i}y^{i+1})\partial_y -
\partial_y(x^{N+1+i}y^{i+1})\partial_x \cr
J^i_N(\alpha) & = \tau^\alpha x^{N+i}y^i,}
$$
with $\tau^\alpha$ representing the generators of SU(N).
The function $x^{N+1+i}y^{i+1}$ will be considered the ``Hamiltonian'' for the
vector field $W^i_N$.  Note that it is sufficient to compute the Poisson
bracket
of the Hamiltonians to get the algebra for the Diff $R^2$ sector.  The use
of Poisson brackets will facilitate our computations in a few moments.
With this representation we will promote our coordinates $x$ and $y$ to
complex coordinates to facilitate the construction of the two cocycle.
In other words we will simply use $\oint dz {z^{p-1}\over 2\pi} =
\delta_{p,0}$.
Let $f$ and $g$ be two arbitrary generators in the $w_\infty$ sector.
Then we may write the two cocycle as
$$
C_{W_\infty}(f,g) = {-c\over 48 \pi^2} \oint dx dy {\partial_x \partial_y f
\partial_x^2
\partial_y g \over y}.
$$
Similarly for $\rho$ and $\gamma$ as arbitrary generators of the
SU(N) sector we have
$$
C_{SU(N)}(\rho,\gamma) = {-k\over 4\pi^2} \oint dx dy {\partial_x \rho \gamma
\over y}.$$
For the coadjoint vectors we will use the notation ${\tilde W^i_M}$ and
${\tilde J^i_M}(\alpha)$.
A pairing between the adjoint vectors and coadjoint vectors in the $w_\infty$
sector is
$$\langle  {\tilde W^i_M} \mid W^j_N \rangle = {-1\over 4\pi^2}\oint dx dy
{\tilde W^i_M} W^j_N = \delta^{i,j}\delta_{M,N},$$
so explicitly we find that $x^{-(M+i+2)}y^{-(i+2)}$ is the dual Hamiltonian
for ${\tilde W^i_M}$
and for the Kac-Moody sector,
$$\langle {\tilde J^i_M(\alpha)} \mid J^j_N(\beta) \rangle =
{-1\over4\pi^2}\oint
dx dy Tr ({\tilde J^i_M(\alpha)}  J^j_N(\beta)) = \delta^{i,j}\delta_{M,N}
\delta^{\alpha\beta},
$$
so that our coadjoint vectors are ${\tilde J^j_N(\alpha)} = \tau^\alpha
x^{-(N+j+1)}y^{-(j+1)}$
where $j \geq 0$.

Now we need to compute the coadjoint action of the group.
Let $(a, \rho, t)$ be an element of the adjoint representation where
the $a$ is the ``Hamiltonian'' of a symplectic diffeomorphism, $\rho$ is valued
in the SU(N) sector, and $t$ is a
central extension.  Using this notation we may write the action of an adjoint
vector on another adjoint vector as
$$
(a, \rho, t) \ast (a', \rho', t') = \big( \{a,a'\}, \{a,\rho'\} + \{\rho, a'\}
+ [ \rho, \rho'],
C_{W_\infty}(a, a') + C_{SU(N)}(\rho, \rho') \big). \eqno(7)
$$
In order to get the adjoint action on a coadjoint vector we consider the
coadjoint vector
$({\tilde b}, {\tilde \gamma}, {\tilde t})$ paired with the adjoint element
$( a', \rho', t')$, i.e., $\langle ({\tilde b}, {\tilde \gamma}, {\tilde t})
\mid
( a', \rho', t') \rangle$.  Then by Leibnitz rule we have that
$$
\big\langle  (a, \rho, t) \ast ({\tilde b}, {\tilde \gamma}, {\tilde t})
\big\vert (a', \rho', t') \big\rangle
+ \big\langle  ({\tilde b}, {\tilde \gamma}, {\tilde t})\big\vert (a, \rho, t)
\ast (a', \rho', t') \big\rangle= 0.
$$
 From Eq.~(7) this implies that
$$
(a, \rho, t) \ast ({\tilde b}, {\tilde \gamma}, {\tilde t}) = \bigg( \{a,
{\tilde b} \} +
\{\rho, {\tilde \gamma} \} -{c {\tilde t} \over 12}{ \partial^3_x \partial_y a
\over y^2},
\{a, {\tilde \gamma} \} + [\rho, {\tilde \gamma}] - k \tilde t {\partial_x \rho
\over y}, 0 \bigg),
\eqno(8)
$$
where we have used the Poisson bracket for all W,W and W,J commutators.

With all of this in place we may now write the action for the coadjoint vector
${\bf B}=(b(x,y),\gamma(x,y), \tilde t)$ as,
$$
\eqalign{
S= {-1\over 4\pi^2}& \int b(x,y) \{ H_\lambda, H_\tau \} d^4x \cr
{-1\over 4\pi^2}&\int {\rm Tr}~ \gamma(x,y) \bigl( \{ H_\lambda,
g^{-1}\partial_\tau g \}
- \{ H_\tau, g^{-1}\partial_\lambda g \} + [g^{-1}\partial_\lambda g,
g^{-1}\partial_\tau g] \bigr) d^4x \cr
{-c \tilde t \over 48\pi^2}&\int {1\over y}( \partial_x \partial_y H_\lambda)
\partial_x^2 \partial_y H_\tau d^4x \cr
{-k \tilde t \over 4\pi^2}& \int {1\over y} \partial_x (g^{-1}\partial_\lambda
g) (g^{-1}\partial_\tau g) d^4x,} \eqno(9)
$$
where the adjoint vectors responsible for transport along the $\lambda$ and
$\tau$ directions
is given by
$$\eqalign{
{\tilde g}^{-1}F^{\tilde g}_\lambda {\tilde g}& = ( H_\lambda, g^{-1}
\partial_\lambda g ,0)\cr
{\tilde g}^{-1}F^{\tilde g}_\tau {\tilde g}& = ( H_\tau , g^{-1} \partial_\tau
g ,0).} \eqno(10)
$$
The Hamiltonians $H_\lambda$ and $H_\tau$ are related to a symplectic
diffeomorphism
of the coordinates $x$ and $y$ by
$ \partial_\rho H_\lambda = {\partial \varphi^\beta \over \partial \lambda}
/{\partial
\varphi^\beta\over \partial x^\alpha} \epsilon_{\alpha\rho}$ and
$ \partial_\rho H_\tau = {\partial \varphi^\beta \over \partial \tau}/
{\partial
\varphi^\beta\over\partial x^\alpha} \epsilon_{\alpha\rho}$,  where the Greek
indices are 1 and 2
and where $\varphi$'s are a two parameter family of symplectic diffeomorphisms
given
by the transformations,
$x \rightarrow \varphi^1(x,y,\lambda,\tau) $  and  $y \rightarrow
\varphi^2(x,y,\lambda,\tau)$.
Using these relations we write the action as
$$
\eqalign{
S= {-1\over 4\pi^2}& \int b(x,y)\biggl( \big({\partial \varphi^\beta \over
\partial \lambda} \big/{\partial
\varphi^\beta\over \partial y}\big) \big({\partial \varphi^\alpha \over
\partial \tau} \big/{\partial
\varphi^\alpha \over \partial x} \big) -
\big({\partial \varphi^\beta \over \partial \tau} \big/{\partial
\varphi^\beta\over \partial y}\big)\big({\partial \varphi^\alpha \over \partial
\lambda} \big/{\partial
\varphi^\alpha \over \partial x}\big) \biggr)  d^4x \cr
{-1\over 4\pi^2}&\int {\rm Tr}~ \gamma(x,y) \biggl( \bigl({\partial
\varphi^\beta \over \partial \lambda} \big/{\partial\varphi^\beta\over \partial
x^\alpha}\bigr)\partial_\alpha (g^{-1}\partial_\tau g)
- \bigl({\partial \varphi^\beta \over \partial \tau}
\big/{\partial\varphi^\beta\over \partial x^\alpha}\bigr)\partial_\alpha
(g^{-1}\partial_\lambda g) \cr
& \quad \quad  +  [g^{-1}\partial_\lambda g, g^{-1}\partial_\tau g] \biggr)
d^4x \cr
\noalign{\vskip5pt}
{-c(k,c_v,26) \tilde t \over 48\pi^2}&\int {1\over y} \partial_x \big({\partial
\varphi^\alpha \over \partial \lambda} \big/{\partial
\varphi^\alpha\over \partial x}\big) \partial_x^2 \big({\partial \varphi^\beta
\over \partial \tau} \big/{\partial
\varphi^\beta\over \partial x}\big) d^4x \cr
{-k \tilde t \over 4\pi^2}& \int {1\over y} \partial_x (g^{-1}\partial_\lambda
g) (g^{-1}\partial_\tau g) d^4x.} \eqno(11)
$$
This is the result that we seek.

We may use this action to study four dimensional gravity by considering an
SU(2)
Kac-Moody algebra.  The spin connections will be SU(2) valued
and again we may write Eq.~(11) as a BF theory.  To see this recall that [2,3]
the action corresponding to gravity via Ashekar variables may be written as
$$\eqalign{
S = &{1\over 2}\int_{R^4} {\rm Tr} \Sigma_\wedge F \cr
= & {1\over 2}\int d^3x dt~ \epsilon^{abc}\{ {\rm Tr}( \Sigma_{bc} F_{0a}) +
{\rm Tr} (\Sigma_{0a} F_{bc}) \},}
$$
where $\Sigma$ is an antisymmetric tensor field for the frame fields and
$F$ is the curvature two form for real SU(2) valued spin connections.
By identifying B of the BF theory with $\Sigma$ we may write
$$\eqalign{
B_{0 a} & = \Sigma_{0 a} = \big({\partial \varphi^\beta \over \partial \lambda}
\big/{\partial
\varphi^\beta\over \partial x}\big)\big({\partial \varphi^\alpha \over \partial
\tau} \big/{\partial
\varphi^\alpha \over \partial y}\big) -
\big({\partial \varphi^\beta \over \partial \tau} \big/{\partial
\varphi^\beta\over \partial x}\big)\big({\partial \varphi^\alpha \over \partial
\lambda} \big/{\partial
\varphi^\alpha \over \partial y}\big) \cr
B_{b c} & = \Sigma_{b c} = \gamma(x,y)\cr
F_{0 a} & = \biggl( \bigl({\partial \varphi^\beta \over \partial \lambda}
\big/{\partial\varphi^\beta\over \partial x^\alpha}\bigr)\partial_\alpha
(g^{-1}\partial_\tau g)
- \bigl({\partial \varphi^\beta \over \partial \tau}
\big/{\partial\varphi^\beta\over \partial x^\alpha}\bigr)\partial_\alpha
(g^{-1}\partial_\lambda g)
  +  [g^{-1}\partial_\lambda g, g^{-1}\partial_\tau g] \biggr) \cr
F_{b c} & = b(x,y)}
$$
where we have suppressed explicit SL(2,R) indices and generators.
With this observation, Eq.~(11) can be used to study 4D gravity
of the type described by Ashtekar and
its order $\hbar$ contributions.  From the evidence of the Atiyah-Singer
index theorems, we suspect that one has integrated both spin one-half
and three-half fields coupled to the connection.  This is because the
spin ${1\over 2}$ sector alone admits no anomaly while the
${3\over 2}$ sector does.  This does not necessarily imply that one needs
supergravity, but that the Rarita-Schwinger field is on the same
fundamental footing as the Dirac (Weyl) fermions.   Let us also
emphasize that the gauge group in Eq.~(11) is arbitrary so that we
may study any self-dual gauge theory.

Notice that the last two terms of Eq.~(11) are exactly analogous to the
Polyakov and WZNW terms.  The Polyakov term is just two separate
2D type terms plus a cross term between $\varphi^1$ and $\varphi^2$
contributions [9].  Since the y-integration only serves to
pick up the $W_2$ sector, we may think of these Polyakov terms
as a generalization of the interacting 2D gravities.
 A generalization to this has been computed in [10]
for the $w_\infty$ algebra that includes central extensions for all conformal
fields.  We prefer to leave the action in a form that is quadratic in the frame
fields.
  The very last term is just the WZNW model when
one expands all terms and integrates by parts.  The y-integration
demands that only the $W^2$ sector or, if you wish, the quadratic
differentials corresponding to the space of metrics, pick up
the anomaly.

Other uses of this action would be to identify the gravitational
instantons and monopoles by using the type of analysis that we used
for QCD strings.  By demanding that $\Sigma$ and F be proportional
to each other, we can identify the isotropy group of the coadjoint
vector and extract gravitational instantons.
The isotropy equation, Eq.~(8), may dictate inequivalent
vacua for 4D gravity.  Different isotropy groups correspond to
different semi-classical vacua.  Furthermore different topologies
can be studied by changing the representation of SDiff $R^2$ to
other structures like SDiff $T^2$.  We will focus on these differences
and the relationship to gravitational instantons in a future publication.

We have seen that the method of coadjoint orbits can lead us to order $\hbar$
effects of a field theory with a rich symmetry.  The method seems to indicate
that the path integral can be computed (up to constants) rather painlessly
to arrive at effective actions.  These effective actions may be seen as
a bosonization prescription (since the fermionic degrees of freedom
are replaced by bosonic degrees of freedom), but perhaps they also contain
contributions from the bound states of the fermions.  Indeed the QCD
discussion leads one to think that the string theory is a string theory of
mesons.  The analog of this statement for the 4D gravity sector would
imply that the spin connections and frame fields are bound states
of more primitive fermionic fields.  The covectors may serve as
external fields and we may promote them to dynamical fields by adding an
appropriate kinetic term.   Another point of interest is
 that we were
able to show that the WZNW models and 2D quantum gravity
are quantum corrections to some Chern-Simons
theories in 3D.  A statement similar to this appears at the 4D level
for the BF theories.  In the spirit of using algebraic
techniques to understand functional integrals we will
use the Moyal brackets [22] to get all higher order corrections
to these theories [23].
This is in accord with the spirit of Ref.~[24].  This idea was used in
the frame work of coadjoint orbits to extend $w_\infty$ to $W_\infty$
in Ref.~[10].  There one finds that (correcting errors in Ref.~[10],
for $ H_\lambda$ and $H_\tau$ (see Eq.~10)), the $W_\infty$
effective action for the orbit corresponding to the coadjoint vector
$ {\bf B}=(\tilde B, t^l \tilde \alpha_l )$ is
$$\eqalign{
&S(\tilde B)= \cr
& \sum^\infty_{i=0} t^i {c_i\over 4\pi^2}  \oint\oint\int\int
  D_y^{2(i+1)}\partial^{i+1}_x\big(y^{i+1}H_\tau\big)~D_y^{2i+2}
\partial^{i+2}_x\big(y^{i+2}H_\lambda\big) ~dx~dy~d\lambda~d\tau \cr
& +  \oint\oint\int\int~ \tilde B(x,y)\{H_\tau, H_\lambda\}_{{}_{\hbar}}~
dx~dy~d\lambda~d\tau, \cr}
$$
where the derivative operator $D_y = \partial_y + {x \over y} \partial_x$.

Lastly, these generalizations may further provide insight
into how one extends the theorems of Ref.~[14]
to non-classical groups.  In particular, the 4D BF action
suggests a relationship between maps from $CP^1 \rightarrow $Diff $S^1$
and self-dual Einstein theories, where the 2D action corresponds
to two separate Liouville-Polyakov theories and an interacting
term between them that appears as a WZ type term.
Also the SU(2) case of Eq.~(11) is suitable for studying a field theoretic
version of reducible, self-dual connections on compact four
manifolds by choosing the isotropy group of the {\bf B} field to be U(1).

\vskip22pt

{\bf Acknowledgements}
V.G.J.R would like to thank A.P. Balachandran, S.J. Gates,
D. Karabali, V.P. Nair,
P. Srinivasan for discussion and encouragement.
This work was supported in
part by NSF Grant PHY-9103914.
\vfill\eject
\nopagenumbers
\baselineskip=13pt
\centerline{\bf REFERENCES}
\bigskip
\settabs 1 \columns
\+ [1]  E. Witten, Nucl. Phys. B311 46 (1988/89) \cr
\+ [2]  J. Samuel, Pram\~ana - J. Phys. 28 L429 (1987)\cr
\+ [3]  Ted Jacobson and Lee Smolin, Class. Quant. Grav. 5 583 (1988)\cr
\+ [4]  A.P. Balachandran and P. Teotonio Sobrinho, \cr
\+    {\it The Edge States of
the BF System and the London Equations}\cr
\+      Syracuse Preprint SU-4228-491 \cr
\+ [5] F. Zaccoria, E.C.G. Sudarshan, J.S. Nilsson, N. Mukunda, G. Marmo,\cr
\+ and A.P. Balachandran,  Phys. Rev D27 2327 (1983);\cr

\+ [6] A.P. Balachandran, G. Marmo, B.S. Skagerstam, and A. Stern,\cr
\+ {\it Gauge Symmetries and Fibre Bundles; Applications to Particle Dynamics}
,\cr
\+ Springer-Verlag Berlin (1983)\cr
\+ [7]  B. Rai and V.G.J. Rodgers, Nucl. Phys. B341 119 (1990); \cr
\+       A. Yu Alekseev and S.L. Shatashvili, Nucl. Phys. B323 719 (1989) \cr
\+ [8]  G.W. Delius, P. van Nieuwenhuizen, and V.G.J. Rodgers,\cr
\+ Inter. Jour. of Mod. Physics A5 3943 (1990) \cr
\+ [9 ]  V.G.J Rodgers,  Mod. Phys. Lett. A 6 1045 (1991) \cr
\+ [10]  V.G.J. Rodgers, Mod. Phys. Lett. A 6 1893 (1991) \cr
\+ [11]  D. Gepner and E. Witten, Nucl. Phys. B 278 493 (1986) \cr
\+ [12]  D. Birmingham, M. Blau, M. Rakowski, and G. Thompson, Phys. Rep. 209
129 (1991) \cr
\+ [13]  V.G.J. Rodgers, {\it QCD Instantons and 2D Surfaces},\cr
\+       Iowa Preprint UI92-2, to appear in Mod. Phys. Lett. A \cr
\+ [14]  M. Atiyah, Comm. Math. Phys. 93 (1984) 437\cr
\+ [15]  P. Goddard and D. Olive, Int. J. Mod. Phys. A 1 303 (1986) \cr
\+ [16]  Q.H. Park, Phys. Lett. 236 B 429 (1990); Phys. Lett. 238 B 287
 (1990);\cr
\+   Phys. Lett. 257 B 105 (1991)\cr
\+ [17]  K. Yamagishi and G.F. Chapline, Class. Quantum Grav. 8 427 (1991)\cr
\+ [18]  K. Yamagishi, Phys. Lett. B 259 436 (1991) \cr
\+ [19]  H. Ooguri and C. Vafa,  Mod. Phys. Lett. A5 1389 (1990);\cr
\+      Nucl. Phys. B361 469 (1991); Nucl. Phys. B367 83 (1991) \cr
\+ [20]  E. Sezgin, {\it Area-Preserving Diffeomorphisms, $w_\infty$ Algebras
and $w_\infty$ Gravity} \cr
\+      Trieste Summer School Lecture, July 17 - Aug. 9, 1991 \cr
\+ [21]  Abhay Ashtekar, `` Non-Perturbative Canonical Gravity,''
World Scientific (1991) \cr
\+ [22]  J. Moyal, Proc. Comb. Phil. Soc. 45 99 (1949) \cr
\+  I. Bakas,  Comm. Math. Phys. 134 487 (1990), Phys. Lett. B228 57 (1989) \cr
\+ [23] R.P. Lano and V.G.J. Rodgers, in progress\cr
\+ [24] E. Bergshoeff, E. Sezgin, Y. Tanii and P.K. Townsend, Ann. Phys.
 199 340 (1990) \cr

\end